\documentclass[iop,deluxetable,twocolappendix]{emulateapj}
\usepackage{amsmath,mathtools}
\newcommand{\Ne}{n_\mathrm{e}}
\newcommand{\td}{\theta_\mathrm{disk}}
\newcommand{\tf}{\theta_\mathrm{funnel}}
\newcommand{\Tf}{T_\mathrm{e,funnel}}
\newcommand{\tb}{{t_\mathrm{begin}}}
\newcommand{\te}{{t_\mathrm{end}}}
\newcommand{\hr}{\mathrm{hr}}
\newcommand{\cm}{\mathrm{cm}}
\newcommand{\mm}{\mathrm{mm}}
\newcommand{\um}{\mu\mathrm{m}}
\newcommand{\Hz}{\mathrm{Hz}}
\newcommand{\sgra}{Sgr~A$^*$}

\begin{document}

\title{Fast Variability and $\mm$/IR flares in GRMHD Models of \sgra\\
  from Strong-Field Gravitational Lensing}

\author{%
  Chi-kwan Chan\altaffilmark{1},
  Dimitrios Psaltis\altaffilmark{1},
  Feryal \"Ozel\altaffilmark{1},
  Lia Medeiros\altaffilmark{2},
  Daniel Marrone\altaffilmark{1},\\
  Aleksander S{\c a}dowski\altaffilmark{3}, and
  Ramesh Narayan\altaffilmark{4}%
}

\email{E-mail: chanc@email.arizona.edu}

\altaffiltext{1}{Steward Observatory and Department of Astronomy,
  University of Arizona,
  933 N. Cherry Ave., Tucson, AZ 85721}

\altaffiltext{2}{Department of Physics,
  Broida Hall, University of California Santa Barbara,
  Santa Barbara, CA 93106}

\altaffiltext{3}{MIT Kavli Institute for Astrophysics and Space Research,
  77 Massachusetts Ave, Cambridge, MA 02139}

\altaffiltext{4}{Institute for Theory and Computation,
  Harvard-Smithsonian Center for Astrophysics,
  60 Garden Street, Cambridge, MA 02138}

\begin{abstract}
  We explore the variability properties of long, high cadence GRMHD
  simulations across the electromagnetic spectrum using an efficient,
  GPU-based radiative transfer algorithm.
  We focus on both disk- and jet-dominated simulations with parameters
  that successfully reproduce the time-averaged spectral properties of
  \sgra\ and the size of its image at $1.3\,\mm$.
  We find that the disk-dominated models produce short timescale
  variability with amplitudes and power spectra that closely resemble
  those inferred observationally.
  In contrast, jet-dominated models generate only slow variability, at
  lower flux levels.
  Neither set of models show any X-ray flares, which most likely
  indicate that additional physics, such as particle acceleration
  mechanisms, need to be incorporated into the GRMHD simulations to
  account for them.
  The disk-dominated models show strong, short-lived mm/IR flares,
  with short ($\lesssim 1\,\hr$) time lags between the mm and IR
  wavelengths, that arise from strong-field gravitational lensing of
  magnetic flux tubes near the horizon.
  Such events provide a natural explanation for the observed IR flares
  with no X-ray counterparts.
\end{abstract}


\section{Introduction}

The black hole in the center of the Milky Way, \sgra, is the optimal
laboratory for studying accretion physics at very low mass accretion
rates.
The mass and distance to the black hole are well constrained from
dynamical studies of the orbits of stars in its vicinity \citep[see,
  e.g.,][]{2008ApJ...689.1044G, 2009ApJ...692.1075G} and its
brightness and variability have been studied extensively across the
electromagnetic spectrum during the last decades (see
\citealt{2001Natur.413...45B} and \citealt{2003Natur.425..934G} for
early studies).

Long monitoring observations have revealed that \sgra\ is highly
variable at timescales comparable to the dynamical time near its event
horizon.
In the infrared (IR, e.g., K band), \citet{2009ApJ...691.1021D} found
that the power spectrum of variability can be modeled by a power law
in the 1--15\,min range with an index that varies between $-1.8$ and
$-2.8$.
Similar power-law indices were obtained in other IR bands and at
230\,GHz \citep{2008ApJ...688L..17M, 2009ApJ...706..348Y,
  2014MNRAS.442.2797D}.
Moreover, \citet{2009ApJ...694L..87M} found evidence for a break in
the submillimeter (submm) and IR power spectrum at $\sim 150\,\min$
\citep[see also][]{2014MNRAS.442.2797D}.

The amplitude of variability depends strongly on the wavelength of
observation and changes from $\sim 10\%$ at $1.3\,\mm$ to $\sim 50\%$
at $2\,\um$, and up to $\gtrsim 100\%$ in the X-rays in three hour
intervals \citep[e.g.,][]{2008ApJ...682..373M, 2008A&A...488..549P,
  2009ApJ...691.1021D}.
In the IR, it is debated whether the variability comprises of peaked
flaring events on top of a more quiescent variability \citep[see][and
  references therein]{2011ApJ...728...37D} or if all the observed
variability is due to a continuum of events with a range of amplitudes
and durations \citep{2012ApJS..203...18W}.
Nevertheless, simultaneous IR/X-ray observations of \sgra\ have
established that even though all strong X-ray flares have IR
counterparts, the opposite is not true \citep{2007ApJ...667..900H}.
This suggests that either there are two physical origins for the IR
events or that the physical mechanism responsible for the flares has
two different observational manifestations.

This wealth of observations has fueled substantial progress in
modeling the accretion flow around \sgra\ \citep[see][for a
  review]{2014ARA&A..52..529Y}.
In particular, the combination of general relativistic
magnetohydrodynamic (GRMHD) simulations and relativistic ray tracing
calculations has led to a relative convergence of numerical models
that account for numerous observed characteristics of \sgra, such as
its quiescent broadband spectrum and the size of its mm image
\citep{2009ApJ...706..497M, 2010ApJ...716..504S, 2010ApJ...717.1092D,
  2015ApJ...799....1C}.
However, a consensus has not been reached on the physical mechanisms
responsible for the rich phenomenology of the variability observed
from \sgra\ and, in particular, for the origin of the observed IR and
X-ray flares.

MHD simulations of accretion flows are inherently turbulent and
variable but, as we showed in \citet{2009ApJ...701..521C}, the
excursions in magnetic field and density generated by MHD turbulence
are not large enough to generate the observed large amplitude of
variability, unless additional physics is incorporated.
In \citet{2009ApJ...701..521C}, we investigated the effect of external
perturbations on the accretion flow.
\citet{2010ApJ...725..450D} explored the possible connection between
flares and the regions of potential magnetic reconnection in MHD
simulations.
\citet{2013MNRAS.432.2252D} investigated the possibility of shock
heating in a tilted accretion disk as an alternate avenue to
generating large variability and flares.

In parallel to the numerical simulations, a number of phenomenological
models of the variability have been explored, which involve, e.g.,
orbiting hotspots or sudden injection of non-thermal particles
\citep[e.g.,][]{2001A&A...379L..13M, 2004ApJ...606..894Y,
  2009ApJ...698..676D, 2009A&A...500..935E, 2014MNRAS.441.1005D}.
These models can reproduce some characteristics of the data but their
connection to the turbulence and thermodynamics of the accretion flow
is unclear.

The developments in the algorithms and the computational power in the
last two years allow us to revisit the question of the broadband
\sgra\ variability using long GRMHD simulations, with high cadence ray
tracing to compute observables, and more detailed descriptions of the
thermodynamics of the accretion flow.
We make use of GRMHD simulations of the accretion flow around
\sgra\ performed with \texttt{HARM} \citep{2012MNRAS.426.3241N,
  2013MNRAS.436.3856S} and radiative transfer calculations performed
with the highly efficient GPU based ray-tracing algorithm
\texttt{GRay} \citep{2013ApJ...777...13C}.
In earlier work, we used the combination of these two algorithms to
carry out a comprehensive study of models with a wide range of black
hole spins, magnetic field geometries, plasma physics models, and
geometric parameters at high time, energy, and spatial resolutions
\citep{2015ApJ...799....1C}.
Among these, we identified five simulations with different
characteristics that all reproduce the observed quiescent broadband
spectrum and the $1.3\,\mm$ image size of \sgra.
We use these as our five baseline numerical models.

In this paper, we calculate multi-wavelength lightcurves with a
$10\,GMc^{-3}$ ($\approx 3.5\,\min$ for \sgra) time resolution that
span approximately $10,000\,GMc^{-3}$ ($60\,\hr$) for each of the five
simulations.
We use these lightcurves to compute the amplitude of variability and
the power spectrum over a wide range of wavelengths and to investigate
the correlations and potential lags between different wavelengths.
We find that the disk-dominated (SANE) models reproduce the short
timescale mm and IR variability with amplitudes and power spectra that
are comparable to those inferred observationally, whereas
jet-dominated (MAD) models produce significantly less variability.
We also find that none of the models produce X-ray flares, which most
likely require additional physics, such as particle acceleration, that
is not captured within the simulations.
We trace the origins of the large amplitude IR variability to
gravitational lensing of short-lived, bright flux tubes near the black
hole horizon, which produces achromatic peaks in the lightcurves.
We discuss the prospect of identifying these lensing events with
future imaging experiments at mm and IR wavelengths.

\section{The GRMHD+ray Tracing Simulations}

\begin{figure}
  \includegraphics[width=\columnwidth, trim=18 12 18 12]{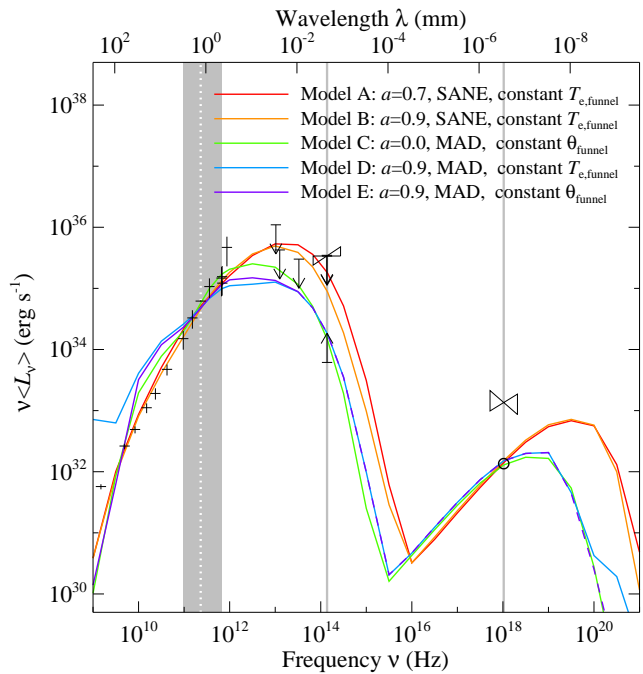}
  \caption{The mean spectra of the five best-fit GRMHD models to
    \sgra\ (see Table~\ref{tab:fit}) and their observational constraints.
    The colorful curves are the mean spectra $\nu\langle L_\nu\rangle$
    computed using all 1024 snapshots.
    The purple curve is plotted in dash for clarity when it overlaps
    with the blue or green curves.
    The crosses, vertical arrows, and bow-ties are the observational
    data that we employed in \citet{2015ApJ...799....1C} to constrain
    the models.
    The gray band between $\nu \approx 10^{11}\,\Hz$ and $\approx
    10^{12}\,\Hz$ marks the frequency range over which we perform a
    least-squares fit to the radio spectrum.
    The gray line at $\nu \approx 10^{14}\,\Hz$ marks the IR frequency
    at which we used the range of fluxes observed at different times
    to impose an upper and lower bound on the models.
    The gray line at $\nu \approx 10^{18}\,\Hz$ marks the X-ray
    frequency, at which we used 10\% of the observed quiescent flux
    (the open circle below the bowtie) to fix the density
    normalization in the flow.
    The white dotted line inside the gray band marks $\lambda =
    1.3\,\mm$, at which we fit the models to the EHT measurement of
    the image size.}
  \label{fig:spec}
\end{figure}

\begin{deluxetable*}{cclccrcrr}
  \tablewidth{6in}
  \tablecaption{Parameters Of The Five GRMHD Models from
    \citet{2015ApJ...799....1C}}
  \tablehead{Model & $a$ & $B_0$ & $\Ne(\td)$ &
    $\td$ & $\Tf$ & $\tf$ & $\tb$\ & $\te$\ }
  \startdata
  A & 0.7 & SANE & $6.9\times10^7$ & 0.02371 &  56.2\ \ \ & ---    & 89,760 &  99,990 \\
  B & 0.9 & SANE & $5.5\times10^7$ & 0.01000 &  31.6\ \ \ & ---    & 39,760 &  49,990 \\
  C & 0.0 & MAD  & $5.9\times10^8$ & 0.00056 & ---\ \ \ \ & 0.0100 & 209,760 & 219,990 \\
  D & 0.9 & MAD  & $1.6\times10^8$ & 0.00075 &   1.8\ \ \ & ---    & 38,520 &  48,750 \\
  E & 0.9 & MAD  & $1.6\times10^8$ & 0.00075 & ---\ \ \ \ & 0.0032 & 38,520 &  48,750
  \enddata
  \tablecomments{The parameters of the five best-fit GRMHD models of
    \sgra\ that we obtained in \citet{2015ApJ...799....1C}.
    The first column lists the model names that we use throughout this
    paper.
    The parameters $a$, $B_0$, $\Ne$, and $\td$ are the black hole
    spin, the initial magnetic field configuration, the normalization
    of the electron number density, and the electron-ion temperature
    ratio in the accretion flow.
    The parameters $\Tf$ and $\tf$ are the dimensionless electron
    temperature and electron-ion temperature ratio in the funnel regions.
    The times listed in the last two columns correspond to the initial
    and final simulation times, in gravitational units, $GMc^{-3}$,
    that we use for the study of variability presented here.
    Note that there are 1024 snapshots for each model separated by
    $10M$, so that $\te - \tb$ is always equal to $10230GMc^{-3}$,
    which is approximately $60\,\hr$ for the mass of \sgra.}
  \label{tab:fit}
\end{deluxetable*}

In \citet{2015ApJ...799....1C} we considered a large number of GRMHD
models, with a wide range of black-hole spins (0 to 0.9), accretion
rates (which determine the scale of electron density in the accretion
flow), initial configurations of the magnetic field (SANE and MAD in
the terminology of \citealt{2012MNRAS.426.3241N}), and with different
prescriptions for the thermodynamic properties of the electrons in the
accretion flow and in the magnetically dominated funnel.
We then used \texttt{GRay} to compute average images and spectra from
100 snapshots from each simulation, with a time resolution of
$10\,GM/c^3$, where $G$ is the gravitational constant, $M = 4.3\times
10^6\,M_\Sun$ is the mass of the Galactic black hole, and $c$ is the
speed of light.
Hereafter, we will use gravitational units with $G = c = 1$.

We identified five GRMHD models that are able to account for the
average spectrum of \sgra\ and the measured size of its emission
region at $1.3\,\mm$, as measured by the EHT.
In particular, we required that each successful model reproduces
{\em (a)\/} the correct average flux and spectral slope at
$10^{11}$--$10^{12}\,\Hz$,
{\em (b)\/} a flux at $\simeq 10^{14}\,\Hz$ that falls within the
observed range of highly variable IR fluxes,
{\em (c)\/} an X-ray flux equal to 10\% of the observed quiescent flux
from \sgra, which has been attributed by \citet{2013ApJ...774...42N}
to emission from the inner accretion flow, and
{\em (d)\/} a scatter-broadened image at $1.3\,\mm$ with a size that is
equal to the characteristic size measured by the EHT
\citet{2008Natur.455...78D}.
We list the parameters of these five models in Table~\ref{tab:fit}.

In this paper, in order to study the variability predicted by the
GRMHD simulations, we use 1024 late-time snapshots of the best-fit
models, with the same time resolution ($10M$).
For the mass of \sgra, this corresponds to a time resolution of
$\approx 212$\,s with a span of $\approx 60\,\hr$.
For each snapshot, we compute four levels of images at each
wavelength, with sizes equal to $1024M \times 1024M$, $256 M \times
256 M$, $64 M \times 64 M$, and $16 M \times 16 M$.
Each level contains $512 \times 512$ pixels.
Hence, the resolution improves by a factor of 4 in each direction
going from one level to the next.
We then calculate the total flux at each time and at each wavelength
by composing together all four levels of images.
In presenting the resulting flux, we often use the quantity $\nu
L_\nu\equiv 4\pi D^2 \nu F_\nu$, which would have been equal to the
luminosity of the source, had the emission been isotropic.

In the ray tracing calculations, we use the \emph{fast light
  approximation}, where the speed of light is assumed to be infinite,
as was done in numerous previous studies \citep{2009ApJS..184..387D,
  2011ApJ...735....9M, 2012ApJ...744..187S}.
This approximation is valid when the light crossing time
$t_\mathrm{crossing} = R/c$ across a region of size $R$ is much less
than the dynamical time scale at that radius $t_\mathrm{dynamical} =
(R^3/GM)^{1/2}$, i.e., when $R \gg GM/c^2$.
This is true for the emission from \sgra\ for wavelengths longer than
$\sim$ mm or shorter than $\sim \um$ (IR).
At mm to IR wavelengths, where the emitting region has a size that is
comparable to the horizon of \sgra, the light crossing time is
$\approx 20$\,s, which is a factor of ten smaller than the time
resolution of the snapshots.

In Figure~\ref{fig:spec}, we plot the average predicted spectrum for
each model together with the observational constraints used in
\citet{2015ApJ...799....1C}.
Even though we have expanded the time span of the snapshots used by a
factor of 10, the average spectra do not violate any of the
observational constraints.

\section{The Variability of Simulated Lightcurves}

\begin{figure*}
  \includegraphics[width=\textwidth]{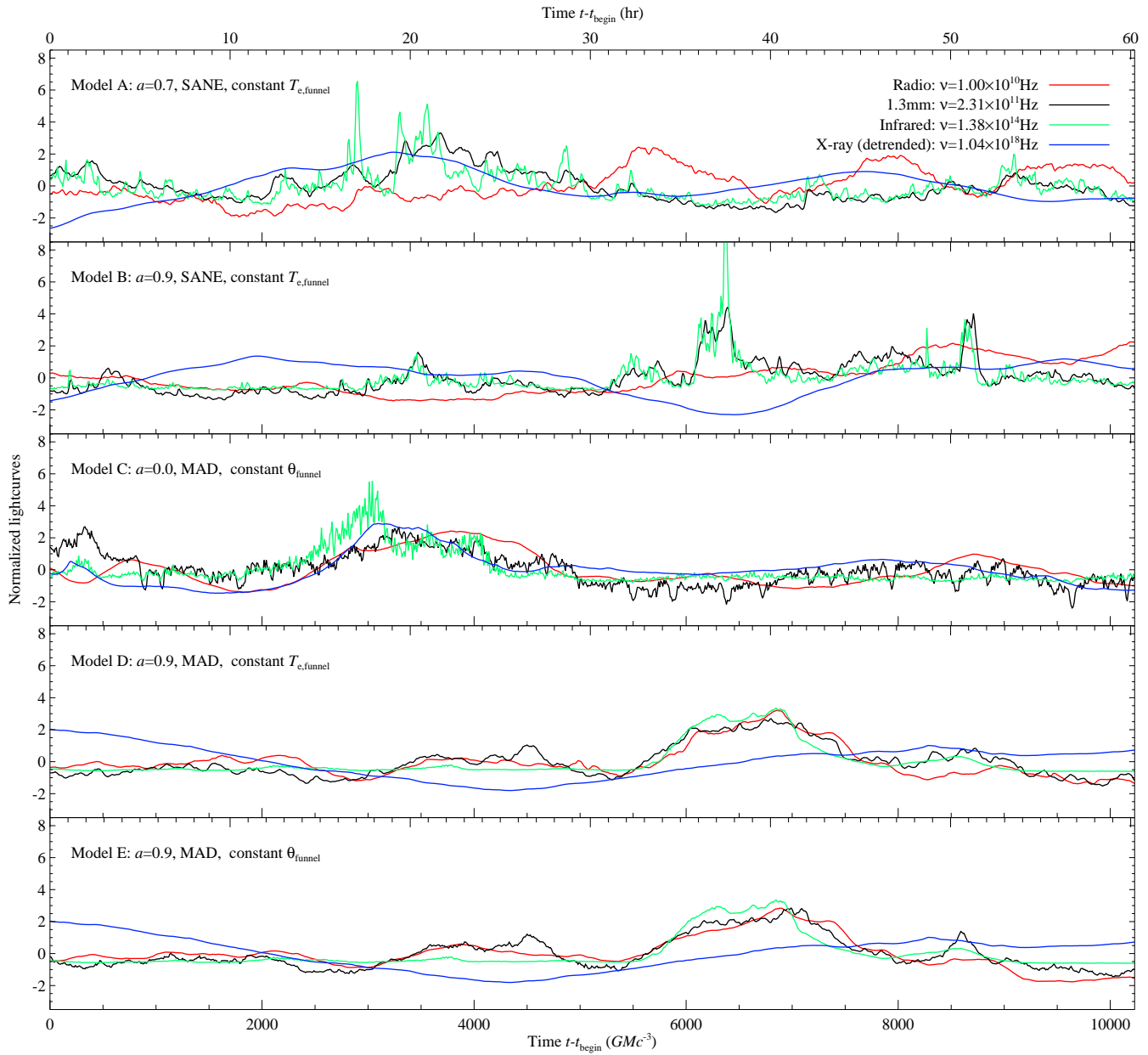}
  \caption{Lightcurves of the best-fit models.
    In each panel, we plot the radio, $1.3\,\mm$, IR, and X-ray
    lightcurves in red, black, green, and blue, respectively.
    We offset and normalize all lightcurves for easy comparison.
    In addition, we also detrend the X-ray lightcurves.}
  \label{fig:lightcurve}
\end{figure*}

\begin{figure}
  \includegraphics[width=\columnwidth, trim=18 12 18 12]{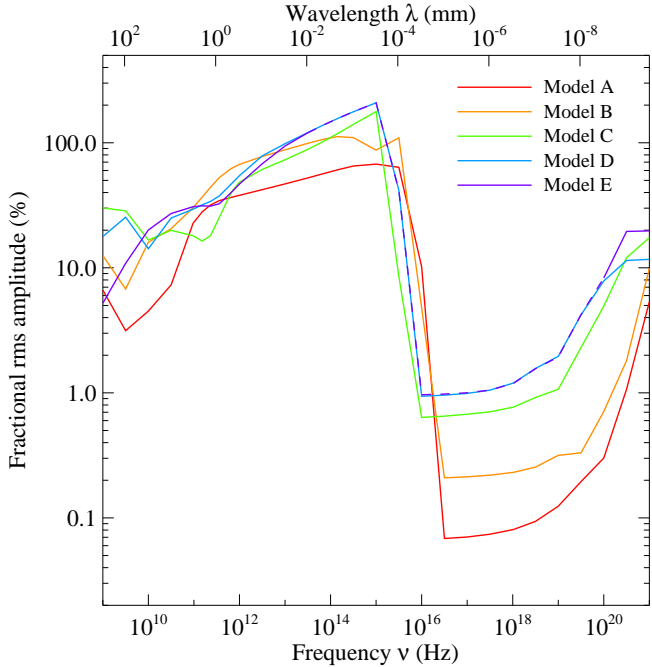}
  \caption{The fractional rms amplitude computed using the full
    lightcurve.
    The color scheme of the amplitudes is identical to the one used in
    Figure~\ref{fig:spec}, except that the purple curve is plotted in
    dash when it overlaps with the blue curves.
    The amplitudes peak near the visible frequencies, i.e., $\nu \sim
    10^{15}\,\Hz$, which is the transition region between the
    synchrotron- and bremsstrahlung-dominated regions.}
  \label{fig:frac}
\end{figure}

\begin{figure*}
  \includegraphics[width=\textwidth, trim=0 6 0 0]{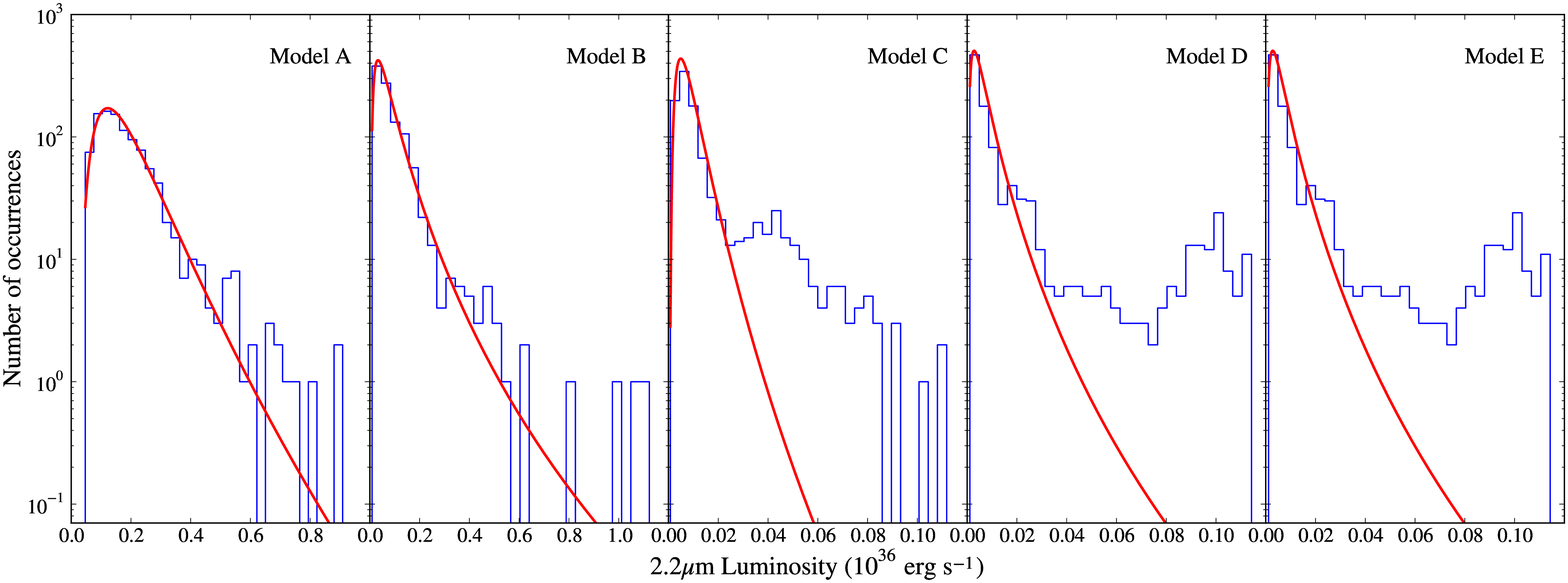}
  \caption{IR flux histograms for the five models, evaluated at
    $2.2\,\um$ (the green curves in Figure~\ref{fig:lightcurve}),
    together with the best-fit log-normal distributions.
    The disk-dominated SANE Models~A and B are well described by the
    log-normal distribution, with marginal evidence for a small excess
    at large fluxes.
    Noting the different scales in the horizontal axes, the
    jet-dominated MAD Models~C--E span a much narrower range in fluxes
    but their distribution is significantly flatter than the
    log-normal curve.}
  \label{fig:hist}
\end{figure*}

In this section, we study the lightcurves from each of the five models
using a variety of time-series analysis techniques.
We compute the power spectra, the fractional rms amplitudes as a
function of photon frequency and investigate correlations between
different frequencies.

\subsection{Lightcurves}

We first compute the lightcurves $L_\nu(t)$ from each of the five
models at four representative frequencies.
In each panel of Figure~\ref{fig:lightcurve}, we plot the radio ($\nu
= 10^{10}\,\Hz$), $1.3\,\mm$ ($\nu=2.31\times10^{11}\,\Hz$), IR ($\nu
= 1.38\times10^{14}\,\Hz$), and X-ray ($\nu = 1.04\times10^{18}\,\Hz$)
lightcurves in red, black, green, and blue, respectively.
We will use this color scheme when we compare different wavelengths
throughout the paper.

The fluxes and the variability amplitudes span a wide dynamical range
across the wavelengths of interest, as we will discuss below.
In order to be able to compare the lightcurves at different
wavelengths, we first subtract the mean flux at each wavelength from
the corresponding lightcurve and normalize it to the standard
deviation calculated over the entire interval of the simulation.
As a result, the mean is zero and the standard deviation is equal to
unity for the normalized flux at each wavelength.
In the X-rays, there is also an overall reduction of flux with time
because of the depletion of the torus that is feeding the black hole.
(This is pronounced in the X-rays because the emission originates from
a large volume).
To remove this artifact of the simulation setup, we also detrend the
X-ray lightcurve by fitting a line to the X-ray flux as a function of
time and subtracting this from the lightcurve.

We show in Figure~\ref{fig:frac} the fractional rms amplitudes over
the entire length of the simulations as a function of photon frequency
for the five different models, which are the standard deviations that
we used to normalize the lightcurves shown in
Figure~\ref{fig:lightcurve}.
At low photon frequencies, the amplitudes increase with frequency,
peaking near the optical at $\approx 10^{15}\,\Hz$.
This is the transition between the synchrotron and the bremsstrahlung
dominated parts of the spectrum.
At the wavelengths where the synchrotron emission dominates, a small
change in the density, temperature, or the magnetic field can cause a
large change in the flux and, therefore, give rise to a large
amplitude of variability.
In contrast, the Bremsstrahlung emission is optically thin and
originates from large volumes, averaging out fluctuations in density
and temperature.

We can draw several conclusions from Figures~\ref{fig:lightcurve} and
\ref{fig:frac}.
First, the jet-dominated models, which are the MAD models D and E,
overall have larger amplitudes of variability but their radio-to-IR
lightcurves are dominated by smoother, long timescale flux changes.
In contrast, the disk dominated models (SANE Models A and B), have
overall smaller amplitudes of variability but their lightcurves are
characterized by short-lived, high amplitude excursions in their IR
and mm flux.
As we will discuss in more detail below, these excursions have a lot
of the characteristics of the IR flares observed from \sgra.
Model C, in which the long-wavelength emission originates primarily
from the base of the jet, has characteristics of both disk- and
jet-dominated models and shows both long and short timescale
variability.
At $1.3\,\mm$, the Model C lightcurve shows evidence for
quasi-periodic dimming, the origin of which we will explore below.
Finally, in the X-rays, all models show very weak and smooth
variations in the flux, with no evidence for any flaring events such
as those observed from \sgra\ by Chandra.

Our multi-wavelength lightcurves allow us to look for correlations
between the variability at different wavelengths.
In the jet-dominated models, all wavelengths (except X-rays) show
quite similar trends in variability and are roughly synchronized.
There is some weak evidence of the radio lightcurve lagging behind the
$1.3\,\mm$ and the IR (see e.g., the times between 6000\,M and 8000\,M
in the lower panels of Figure~\ref{fig:lightcurve}).
In the disk-dominated models, it is only the $1.3\,\mm$ and the IR
lightcurves that vary in sync, while the radio is decoupled.
Moreover, the sharp IR flares have nearly coincident, albeit weaker,
$1.3\,\mm$ counterparts with only very short ($\lesssim 1\,\hr$) time
lags.
In all models, the X-ray variability is decoupled from the long
wavelengths, which is expected given their very different emission
locations in the flow.

\subsection{Flux Histograms}

The range and character of variability observed from \sgra\ is often
investigated via studies of flux histograms \citep[see,
  e.g.,][]{2006ApJ...644..198Y, 2009ApJ...706..348Y,
  2009ApJ...691.1021D, 2011ApJ...728...37D, 2012ApJS..203...18W}.
Figure~\ref{fig:hist} shows the flux histograms at $2.2\,\um$ for our
five models over the entire simulation time, with each flux
corresponding to a time integration of $10M$.
(We chose this wavelength for this plot, in order to compare it
directly to the analysis of \citealt{2011ApJ...728...37D}.)
As in the case of the observational studies, we fit the low-flux ends
of the histograms (up to $0.6 \times 10^{36}$\,erg\,s$^{-1}$ for
Models~A and B and up to $0.03 \times 10^{36}$\,erg\,s$^{-1}$ for
Models~C--E) with a log-normal distribution.

The flux histograms of the disk-dominated SANE Models~A and B are well
described by the log-normal distribution with marginal evidence for a
small excess at large fluxes.
In other words, the same physical mechanism appears to be responsible
for most of the variability, with some evidence that the brightest of
flares occur at larger than expected fluxes.
As we will discuss in the next section, this is consistent with our
interpretation that most of the variability in these models is the
result of the presence of short-lived hot, magnetically dominated flux
tubes in the accretion flow.
However, the brightest flares are generated when one of these
magnetically dominated regions gets lensed behind the black hole and
its brightness is greatly magnified.

The flux histograms of the jet-dominate MAD Models~C--E are instead
significantly flatter than the log-normal distribution at all but the
smallest fluxes.
This is another illustration of the fact that the IR lightcurves of
these models show slow variability over a range of fluxes that cannot
be described in terms of a range of short-lived flares.

\subsection{Fractional RMS Amplitudes}

In Figure~\ref{fig:frac}, we looked at the frequency dependence of the
fractional rms amplitude using the entire $\sim 60\,\hr$ time span of
the simulations.
However, given the continuum of flux excursions and timescales visible
in Figure~\ref{fig:lightcurve}, these amplitudes will depend on the
time span used to calculate the rms and the particular interval for
which it is computed.
This is especially important when comparing theoretical amplitudes to
observations, which usually span intervals that are much shorter.

\begin{figure*}
  \includegraphics[width=\textwidth, trim=0 6 0 0]{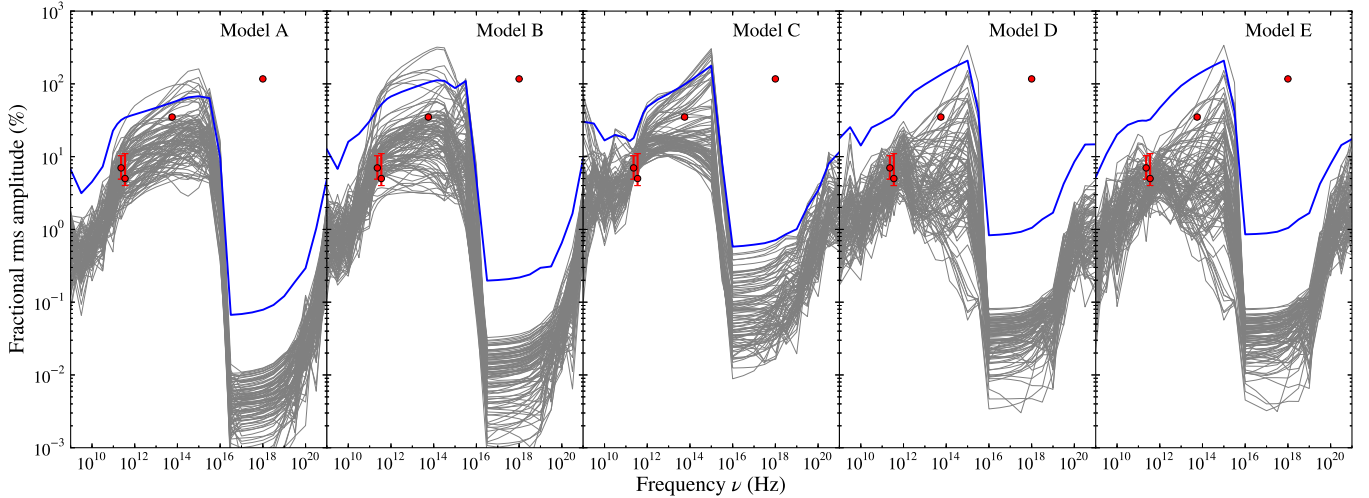}
  \caption{Gray curves show the fractional rms amplitudes computed
    using numerous overlapping $510 M$ ($\approx 3\,\hr$) intervals
    from the simulations, starting at times $t-\tb = 0, 100 M, 200 M, \dots$.
    The blue curve in each panel, which shows the rms amplitude as a
    function of photon frequency computed using the entire span of the
    simulation, is identical to the corresponding curve in
    Figure~\ref{fig:frac}.
    The red data points and error bars mark the range of rms
    amplitudes measured over a number of overlapping 3\,hr intervals
    during the 2006 July 17 observations reported in
    \citet{2008ApJ...682..373M}.}
  \label{fig:frac_short}
\end{figure*}

In order to carry out the comparison to observations, we use the
simultaneous multi-wavelength monitoring of \sgra\ of
\citet{2008ApJ...682..373M}, which is one of the few data sets that
spans the wavelengths from radio to X-rays.
In particular, we use the 2006 July 17 lightcurve, during which an
X-ray flare was detected.
The longest time span of simultaneous observations is equal to 3\,hr
and the range of rms amplitudes measured over a number of overlapping
3\,hr intervals during this observation is shown as red data points
with error bars in Figure~\ref{fig:frac_short}.
In the same figure, we plot as gray lines the predicted fractional rms
amplitudes computed from many overlapping $510\,M$ ($\approx 3\,\hr$)
intervals from the simulations.

As expected, the simulations show a wide range of fractional
amplitudes at each wavelength depending on the particular realization
of the turbulent magnetic field and density in the accretion flow.
However, all realizations have the same overall wavelength dependence
and peak at optical frequencies.
Furthermore, the data points in the mm and IR overlap with typical
realizations of the simulated curves.
On the other hand, the X-ray variability amplitude, which reflects the
presence of the X-ray flare, is many orders of magnitudes larger than
any of the predicted realizations.

This comparison between the observed and simulated variability
amplitudes indicates that GRMHD simulations naturally produce IR
flares and variability with no X-ray counterparts.
Indeed, observations show two types of flaring events; IR flares are
often not accompanied by X-ray counterparts, whereas X-ray flares
always have IR counterparts \citep{2007ApJ...667..900H}.
Interpreting the observations in the context of our simulations
suggests that there are two distinct mechanisms that cause the two
types of flaring activity.
As we will discuss in detail in the next section, strong-field
gravitational lensing in the turbulent GRMHD flows naturally produces
the majority of IR flares without X-ray counterparts.
However, additional physical processes must be incorporated into the
simulations to generate the X-ray flares.

\begin{figure*}
  \begin{center}
    \includegraphics[width=\textwidth, trim=0 18 0 -18]{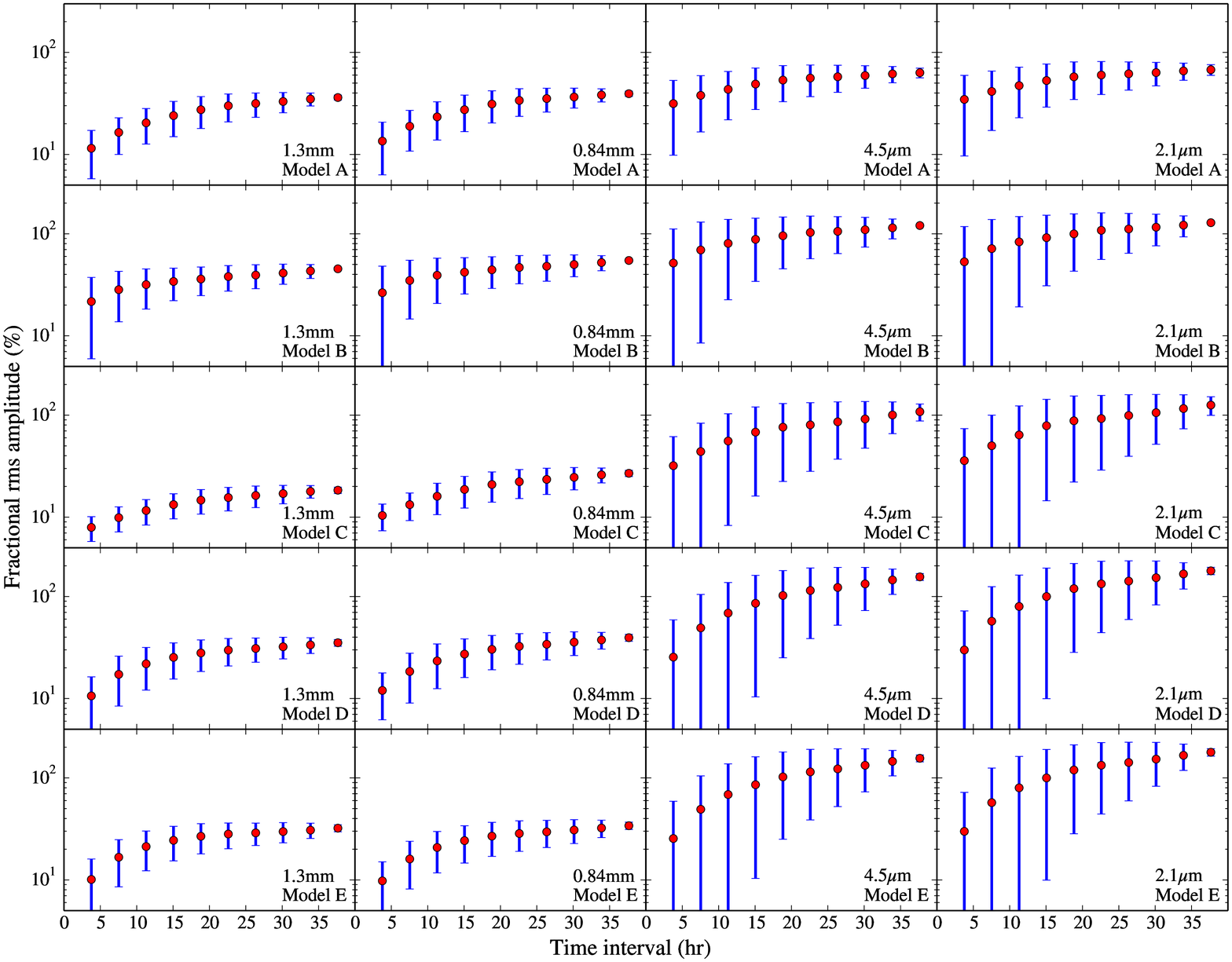}
  \end{center}
  \caption{The fractional rms amplitude at $1.3\,\mm$, $0.86\,\mm$,
    $4.5\,\um$, and $2.1\,\um$, as a function of the time interval
    over which it is evaluated.
    For each value of the time interval, the red points correspond to
    the average fractional rms amplitude for all the overlapping
    segments in the simulation starting at times $t-\tb = 0, 100 M,
    200 M, \dots$ and the error bars represent the variance among them.}
  \label{fig:struct}
\end{figure*}

Given the sensitivity of the variability amplitude on the time span
used, we can use the simulations to explore this dependence in detail.
In Figure~\ref{fig:struct}, we show the fractional rms amplitude at
two different mm and IR wavelengths each, as a function of the time
interval over which it is evaluated.
For each value of the time interval, the red points correspond to the
average fractional rms amplitude for the same overlapping segments in
the simulation discussed earlier and the error bars represent the
variance among them.
As expected, the overall trend is that the amplitude increases with
increasing time interval, with no evidence for a characteristic time
scale in the range explored (i.e., 5--40\,hr).
This is typical of red noise processes, which we can explore in more
detail using power spectral techniques.
In addition, the fact that the fractional rms amplitude in the IR
becomes larger than 100\% points to a large number of short-lived,
high amplitude flux excursions (flares) being responsible for this
variability.
This observation also explains the fact that the range of variability
amplitudes in different 3\,hr segments is larger in the IR than at mm
wavelengths, since the amplitude of any particular segment in the IR
depends strongly on whether a strong flare has occurred within its
duration.

\subsection{Power Spectra}

Red noise variability is best studied in Fourier space, using
power-spectral techniques.
However, generating a power spectrum requires uniformly sampled
lightcurves, with fast sampling and minimal data gaps.
This is very difficult to achieve with ground-based observations of
\sgra\ (e.g., in the mm and in the IR), which would require data
spanning many hours but sampled every few minutes.
For this reason, very few attempts have been made to measure the power
spectrum of \sgra, and the few efforts have primarily been in the IR
via the measurement of the structure function.
In all cases, the power-spectrum was modeled by a power-law function
(\citealt{2008ApJ...688L..17M}, \citealt{2009ApJ...691.1021D},
\citealt{2010A&A...510A...3Z}, and \citealt{2009ApJ...706..348Y} for
near-infrared, and \citealt{2014MNRAS.442.2797D} for submm).
In a very detailed study of the systematic uncertainties that are
inherent to the measurement of power spectra via structure functions,
\cite{2009ApJ...691.1021D} reported indices for the K~band power
spectra in the range $-1.8$ to $-2.8$, while
\citet{2008ApJ...688L..17M} reported evidence for a break in the power
spectrum at $154^{+124}_{-87}$\,minutes, with the power-law index
changing from $-2.1\pm 0.5$ above this break to $-0.3^{+0.2}_{-0.4}$
below it (see also \citealt{2014MNRAS.442.2797D} for a similar report
of a break at $8^{+3}_{-4}\,\hr$ for submm wavelengths).

\begin{figure*}
  \includegraphics[width=\textwidth]{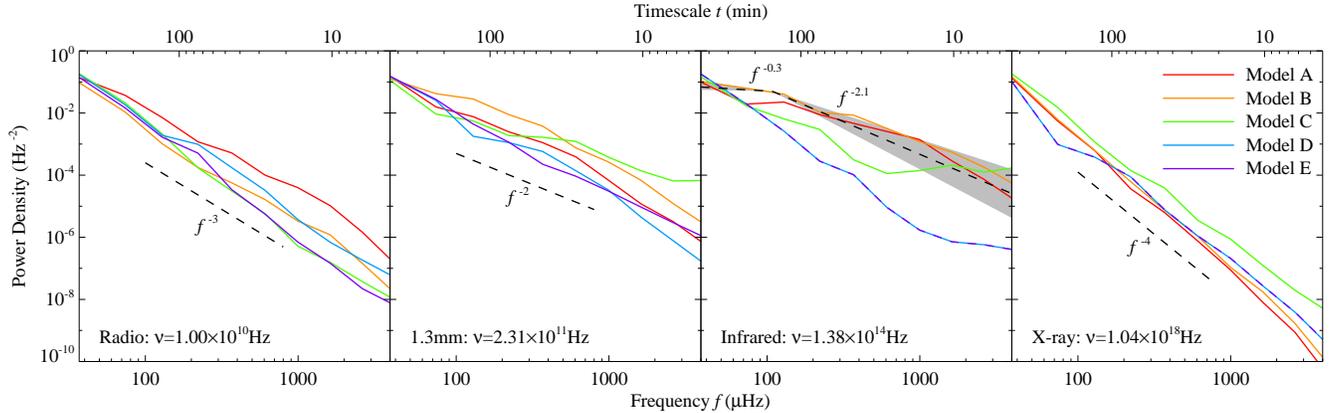}
  \caption{Power density spectra of the normalized lightcurves shown
    in Figure~\ref{fig:lightcurve}.
    In each panel, five power spectra are plotted, one each for
    Models~A--E.
    The panels, from left to right, are the power spectra of the
    lightcurves in the radio, $1.3\,\mm$, IR, and X-rays.
    The color scheme matches the one we used in Figure~\ref{fig:spec},
    with the exception that, for the IR and X-rays (the third and
    fourth panels), we plot the purple curves in dash as they overlap
    with the blue curves.
    In the third panel, the shape of the power spectrum inferred
    observationally by \citet{2008ApJ...688L..17M} is shown as a
    dashed black curve, with its uncertainty depicted by a shaded gray band.
    In the remaining three panels, the dashed lines show power-law
    dependencies that are similar to those exhibited by the simulation
    data.}
  \label{fig:pspec}
\end{figure*}

To make a quantitative comparison of the variability predicted in the
GRMHD simulations to the variability reported in these studies, we
compute the power spectra for each of the five models at four
different photon frequencies and plot them in Figure~\ref{fig:pspec}.
To compute the power spectra, we divide the 1024 snapshots from each
lightcurve into 7 overlapping segments, each of which contains 256
data points.
We multiply these segments by a symmetric ``triangle'' window
function.
We then compute the fast Fourier transform of the windowed segments
and average them.
Finally, we bin the spectra with a bin size equal to the Fibonacci
sequence.
Each panel of Figure~\ref{fig:pspec} contains the power spectra for
the five models A--E, color-coded as before, for the radio,
$1.3\,\mm$, IR, and the X-ray lightcurves.

The power spectra in the X-rays (fourth panel in
Figure~\ref{fig:pspec}) show an $f^{-4}$ dependence, which is
characteristic of long-wavelength variability over timescales that are
longer than the span of the simulations.
We can, therefore, use them only to infer that there is no evidence
for fast X-ray variability in the simulations.
In all three of the longer wavelengths, however, the models give rise
to power spectra with a range of flatter slopes.
In the radio, the slopes can be approximately characterized as
$f^{-3}$, with Models~C and E having somewhat steeper power spectra.
At $1.3\,\mm$, all models show a flatter dependence, with slopes
comparable to $\approx -2$.
In addition, Model~C has the flattest spectrum among all of the
models, consistent with the fact that it has the strongest
short-timescale variability at this wavelength (see
Figure~\ref{fig:lightcurve}).

In the IR, the two disk-dominated models (A and B) produce higher
relative power at higher frequencies, as expected from the visual
inspection of lightcurves that show high levels of short timescale
variability.
These two models also exhibit some weak evidence for flattening toward
smaller frequencies.
To compare these power spectra to observations, we also plot in dashed
line a power spectrum that has a break at 154\,min, with slopes $-0.3$
and $-2.1$ below and above the break and a gray band that corresponds
to the formal uncertainties in these slopes \citep[as inferred
  by][]{2009ApJ...694L..87M}.
This figure demonstrates that not only the overall amplitude but also
the detailed power spectrum of variability generated by our
disk-dominated GRMHD simulations are consistent with
\sgra\ observations.
Therefore, the observed IR power spectra appear to favor the
disk-dominated SANE models as opposed to the jet-dominated MAD ones.

\subsection{Correlations}
\label{sec:correlations}

A higher order Fourier statistic that is often used in the variability
studies of \sgra\ is the cross correlation function between different
wavelengths that is used to search for time lags in the lightcurves.
Such lags are often interpreted as signatures of short-lived, hot
emitted regions that are advected with the accretion flow or are
ejected in a jet or outflow \citep[e.g.][]{2008ApJ...682..373M,
  2009A&A...500..935E, 2015A&A...576A..41B}.
We compute the cross correlation between the $1.3\,\mm$ lightcurve and
the lightcurves at other wavelengths to look for such signatures.

\begin{figure*}
  \includegraphics[width=\textwidth]{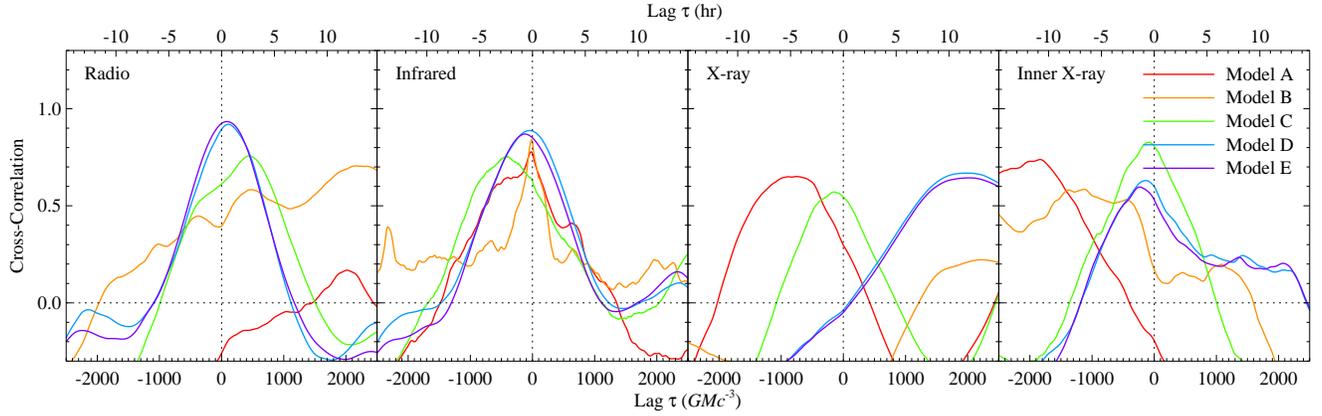}
  \caption{Cross correlations between the lightcurves at different
    wavelengths and the lightcurve at $1.3\,\mm$, for the five models
    shown in Figure~\ref{fig:lightcurve}.
    For the last panel (labeled as inner X-rays), we consider only the
    emission that originates from the inner $16M\times16M$ portion of
    each image.}
  \label{fig:ccorr}
\end{figure*}

The first three panels in Figure~\ref{fig:ccorr}, from left to right,
show the cross correlations as functions of time lag between the
$1.3\,\mm$ lightcurve and the radio, IR, and X-ray lightcurves,
respectively.
A positive lag implies that the lightcurve at the particular
wavelength lags behind the $1.3\,\mm$ lightcurve, whereas a negative
lag implies that it leads the $1.3\,\mm$ lightcurve.
There are clear differences between the $1.3\,\mm$--radio correlations
and $1.3\,\mm$--IR correlations with the $1.3\,\mm$--X-ray
correlations.
This is not surprising, because most of the X-ray emission comes from
large radii, while the $1.3\,\mm$ emission originates near the event
horizon.
In order to focus on the $1.3\,\mm$--X-ray correlations of the inner
flow, we also compute X-ray lightcurves using the inner $16M\times16M$
images only, and cross correlate the $1.3\,\mm$ lightcurves with these
inner X-ray lightcurves.
The result is plotted in the fourth panel of Figure~\ref{fig:ccorr}.

In the disk-dominated, SANE, models, there is clear evidence for
correlated variability only between $1.3\,\mm$ and IR curves with
small ($\lesssim 1\,\hr$) time lags.
In Section~\ref{sec:lensing} we use the simulated images to argue that
variability due to strong-field gravitational lensing is the cause for
the achromatic character of the large flares seen in the simulations
as well as for the lack of time lags.
On the other hand, there is no correlated variability between radio,
$1.3\,\mm$, and X-ray wavelengths, as can also be seen in the
lightcurves shown in Figure~\ref{fig:lightcurve}.
The two peaks appearing at time lags $\sim 10\,\hr$ for models A and B
between the radio and $1.3\,\mm$ as well as the two peaks at $\sim
-10\,\hr$ time lags between the inner X-rays and the $1.3\,\mm$ simply
correspond to the time differences between physically unrelated peaks
in the lightcurves.
Such chance events emphasizes the difficulty of using
cross-correlation analysis in short observations to infer the time
evolution of emission structures in the accretion flows.

In the jet-dominated, MAD, solutions, the radio, $1.3\,\mm$, and IR
lightcurves are highly correlated, with marginal evidence of radio
lightcurves lagging and the IR leading the $1.3\,\mm$ lightcurves by
$\sim 1\,\hr$.
There also seems to be a similar level of correlation between the
inner X-ray and $1.3\,\mm$ lightcurves, with the former leading the
latter by $\sim 1\,\hr$.
This ordering suggests that the physical mechanism driving the
variability in these solutions causes first the X-rays (in the inner
disk), then the IR, then the $1.3\,\mm$, then finally the radio flux
to increase.
In section~\ref{sec:reconnection}, by studying the simulated images,
we conjecture that magnetic reconnection is responsible for the
variability in these models.

\begin{figure*}
  \includegraphics[width=\textwidth]{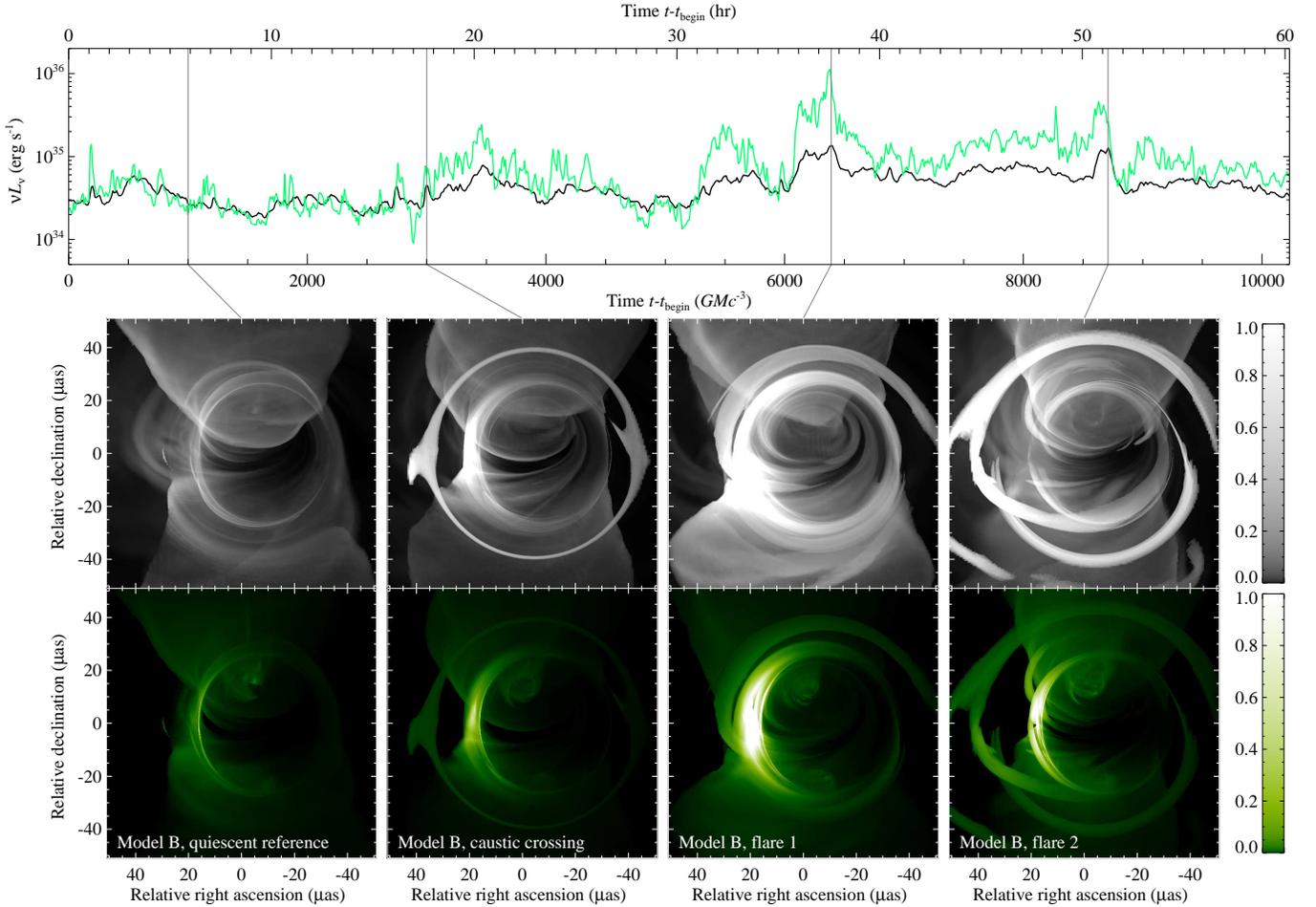}
  \caption{The $1.3\,\mm$ {\em (black)\/} and IR {\em (green\/)}
    lightcurves for Model~B (neither rescaled nor normalized).
    The middle and lower rows show the $20M\times 20M$ images at
    $1.3\,\mm$ and in the IR at different points in time that
    correspond to a quiescent reference and to several flares.
    All images use the same quasi-logarithmic color scale (see
    adjacent color bars), where unity represents the minimum of the
    peak brightness value of the images in each row.
    The small-amplitude flare in the second column is generated by
    strong-field gravitational lensing of a small, magnetically
    dominated region that crossed a caustic behind the black hole and
    gave rise to a nearly perfect Einstein ring.
    The large-amplitude flares in the third and fourth columns are
    generated by sheared magnetically dominated regions, the light
    from which appears both above and below the equatorial plane,
    because of gravitational lensing.}
  \label{fig:filament}
\end{figure*}

\begin{figure*}
  \includegraphics[width=\textwidth]{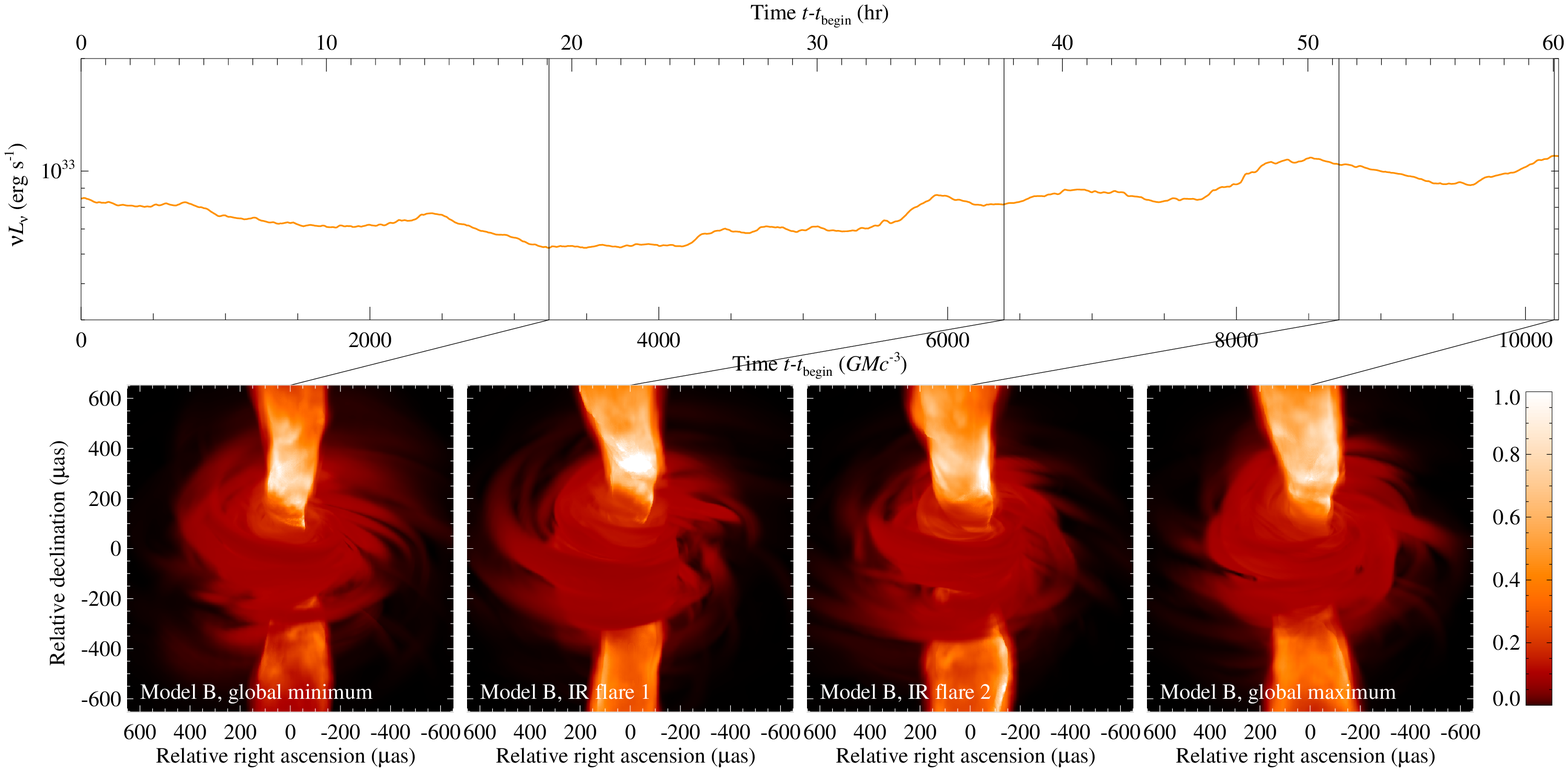}
  \caption{The radio (3\,cm) lightcurve for Model~B (neither rescaled
    nor normalized) and the $256M\times 256M$ images at different
    points in time corresponding to the global minimum of the
    lightcurve, the two flare events shown in
    Figure~\ref{fig:filament}, and the global maximum.
    All images use the same quasi-logarithmic color scale (see
    adjacent color bar), where unity represents the minimum of the
    peak brightness value of all images.
    The emission originates primarily in the hot, magnetically
    dominated funnel and is partially obscured by the colder accretion
    flow.
    Any gravitationally lensed structures, such as those shown in
    Figure~\ref{fig:filament} for the $1.3\,\mm$ and IR images, are
    completely obscured by the cold accretion flow.}
  \label{fig:obscure}
\end{figure*}

\section{The Physical Origins of the Variability}

In the previous section, we focused on calculating lightcurves of the
simulated emission at different wavelengths and their statistical
properties.
In this section, we take a deeper look at the simulated time-dependent
images in order to explore and understand the physical origins of the
different types of variability that we see in the simulations.

\subsection{Disk-dominated SANE models}

The variability in the disk-dominated SANE models is characterized by
a spectrum of intense, short lived flares in the IR, with related
(albeit weaker) flux excursions at $1.3\,\mm$, and uncorrelated slow
variability in the radio and in the X-rays.

We first discuss the origin of the IR/$1.3\,\mm$ flares, which we
attribute to strong-field gravitational lensing of hot flux tubes in
the accretion flow and then explore the origin of the radio
variability, which we attribute to the interplay between emission from
a bright funnel and osculation by a cold accretion disk.

\subsubsection{Strong-Field Lensing of Magnetic Flux Tubes}
\label{sec:lensing}

In Figure~\ref{fig:filament} we plot the $1.3\,\mm$ and IR lightcurves
of Model~B as well as the corresponding images at a quiescent
reference and during several representative flares, with different
amplitudes.
It is clear from the achromatic character of the bright structures in
the images as well as from their characteristic shapes that
strong-field gravitational lensing plays the dominant role in
generating these flares.

In some instances, such as the one shown in the second column from the
left in Figure~\ref{fig:filament}, a hot, magnetically dominated
structure that has a small spatial extent crosses a caustic, giving
rise to a nearly perfect Einstein ring and an accompanying flux
increase.
This is the phenomenon that was investigated by
\citet{1994ApJ...421...46R}, who calculated the properties of caustics
around rotating black holes.
In other instances, the lensing of a sheared flux tube of small radial
but larger azimuthal extent produces images of this structure to
appear both above and below the disk and causes large amplitude flares
(see the right two columns of Figure~\ref{fig:filament}).
This phenomenon is very similar to the large increase of the
brightness of stars during microlensing events.

Because of their gravitational origin, the presence of the large flux
excursions in our simulations depends very weakly on the simplifying
assumptions that are inherent to our model.
The only requirement is the fact that the flow needs to be optically
thin in order for the lensed images not to be obscured by intervening
material.
However, the actual spectra of the flares depend strongly on our
prescription for the plasma thermodynamics in the magnetically
dominated regions.
We recall that, in both Models~A and B, when the plasma-$\beta$
parameter exceeds a threshold (here $\beta > 0.2$) we use a thermal
plasma model with a constant electron temperature \citep[see
  discussion in][]{2015ApJ...799....1C}.

The plasmas in these short-lived, magnetically dominated filaments may
not have time to thermalize completely and this will affect the
particular spectrum of the flares, predominantly in the IR.
Nevertheless, within our simplifying assumptions, the IR luminosity at
$2.1\,\um$ becomes an order of magnitude larger than $1.3\,\mm$
luminosity during the strong flares, from which we can infer the
spectral index between these two wavelengths as
\begin{equation}
  \alpha \simeq
  \frac{\log(L_{2.1\,\um}/L_{1.3\,\mm})}{\log(2.1\,\um/1.3\,\mm)}
  \simeq -0.4\;.
\end{equation}
Even though this is not directly comparable to the $\alpha\simeq
-0.6\pm 0.2$ value inferred observationally by
\citet{2007ApJ...667..900H} for the index of the IR spectrum alone, it
is significant that our simple thermal model for the electrons in the
magnetic filaments generate a flare spectrum that is substantially
bluer than that of the average emission.

\subsubsection{Bright Radio Funnels Obscured by\\
  Optically Thick Disks}
\label{sec:obscure}

In section~\ref{sec:correlations}, we asserted that, in the
disk-dominated SANE models, the large ($\gtrsim 10\,\hr$) positive
time lags obtained between the radio and the $1.3\,\mm$ lightcurves
are artifacts of matching physically unrelated flux variations between
these wavelengths.
Figure~\ref{fig:obscure} shows the radio ($3\,\cm$) images calculated
during the global minimum of the lightcurve, the two flare events
shown in Figure~\ref{fig:filament}, and the global maximum of the
lightcurve.
A comparison of Figures~\ref{fig:filament} and \ref{fig:obscure} shows
that the emission at the two wavelengths originates from two
different, and disconnected locations in the flow.
The radio emission comes from a hot funnel, which is partially blocked
by a colder, optically and geometrically thick accretion disk.
Note also that the scale of these images is much larger than the
scales of the $1.3\,\mm$ and the IR images shown in
Figure~\ref{fig:filament}.
In contrast, all gravitationally lensed magnetic filaments that are
responsible for the large $1.3\,\mm$/IR flares lie very close to the
horizon and are completely obscured by the cold accretion disk at
radio wavelengths.

\subsection{Jet-dominated MAD models}

\begin{figure*} 
  \includegraphics[width=\textwidth]{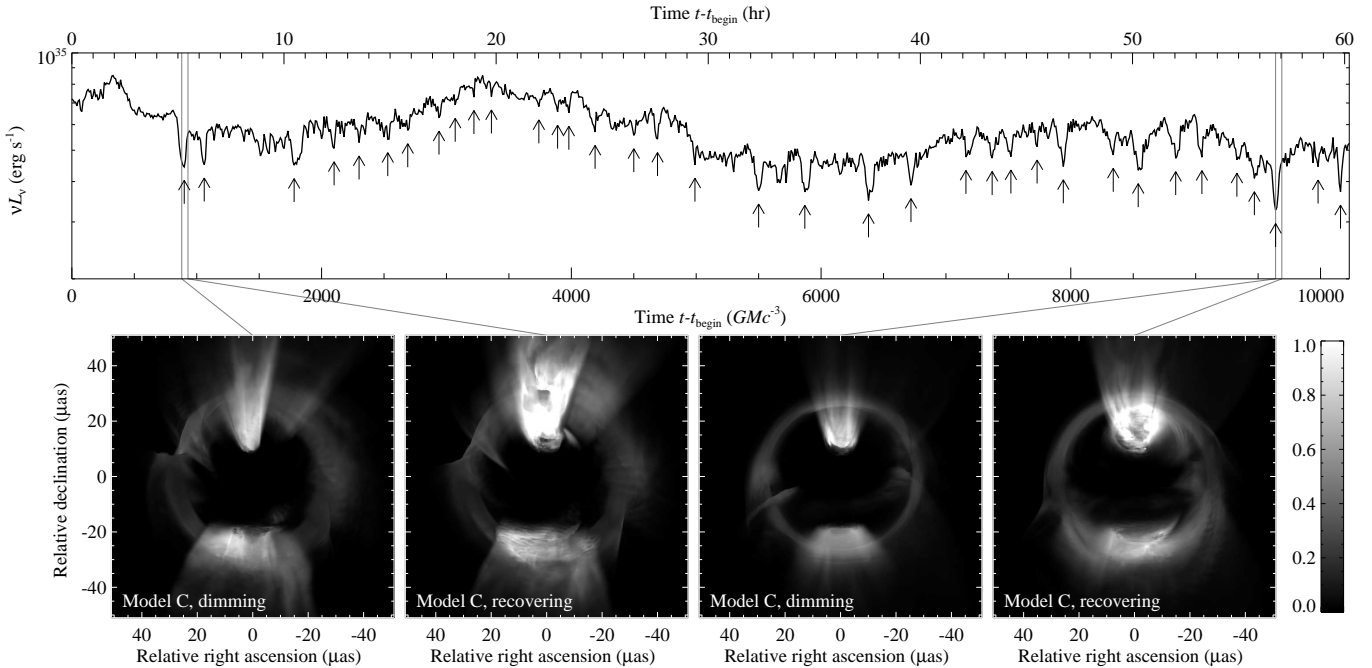}
  \caption{The $1.3\mm$ lightcurve for Model~C and the corresponding
    $20M\times 20M$ images during representative short timescale,
    quasi-regular dimming episodes.
    Because the variation in the flux during each event is small, we
    use a linear color scale for images (see adjacent color bar, where
    unity represents the minimum of the peak brightness values of the images).
    Each dimming event occurs when the mass loading of the footpoint
    of the jet funnel is reduced and the funnel itself becomes narrower.}
  \label{fig:wave}
\end{figure*}

\begin{figure*}
  \includegraphics[width=\textwidth]{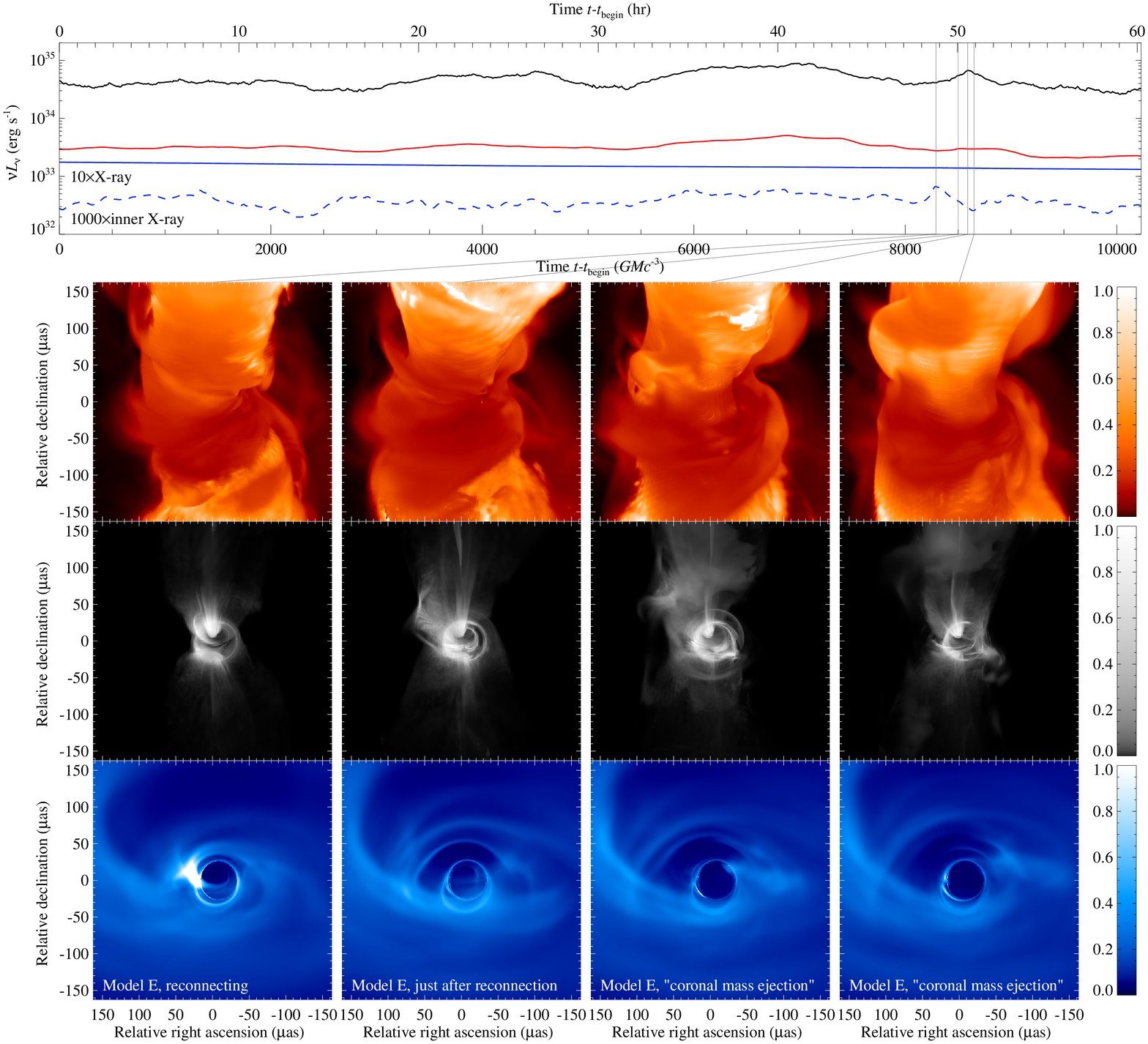}
  \caption{The radio {\em (red)\/}, $1.3\,\mm$ {\em (black)\/}, and
    X-ray {\em (blue)\/} lightcurves for Model~E and the corresponding
    $64M\times 64M$ images after a representative reconnection event.
    The fluxes in the total and inner X-ray lightcurves have been
    scaled by factors of 10 and 1000, respectively, for clarity.
    The color scales of the images is quasi-logarithmic (see adjacent
    color bars, where unity represents the minimum of the peak
    brightness values of the images for each row).
    The first column shows a snapshot where the inner X-ray lightcurve
    reaches a local maximum.
    We attribute this to a reconnection event, where magnetic energy
    is dissipated into thermal energy and the topology of the field
    changes.
    This generates a hot spot in the X-ray image, at the left side of
    the black-hole shadow.
    In the second column, a twisted feature develops in the left hand
    side of the $1.3\,\mm$ image and moves upwards in the third
    column, at which point the $1.3\,\mm$ lightcurve reaches a local maximum.
    Finally, the fourth column shows a snapshot where the inner X-ray
    lightcurve reaches a local minimum, but the first twisted feature
    appears in the funnel in the radio image above the cold, obscuring
    accretion disk.
    This set of events resembles coronal mass ejection episodes
    observed in the sun.}
  \label{fig:recon}
\end{figure*}

The lightcurves of the jet-dominated, MAD models do not show any
evidence for short-lived flaring events but rather slower, correlated
flux variations across the radio-to-IR part of the spectrum.
Moreover, there is marginal evidence for the IR lightcurve leading the
$1.3\,\mm$ one by $\simeq 1\,\hr$ and the $1.3\,\mm$ lightcurve
leading the radio ($3\,\cm$) one by a comparable amount.

In this section, we attribute a set of quasi-regular dimming episodes
in the $1.3\,\mm$ lightcurve of Model~C to a time-dependent loading of
the footpoint of the jet funnel.
We also point out some evidence for variability generated by
dissipation of magnetic energy in localized events.

\subsubsection{Time-variable Loading of the Jet Funnel}

The non-spinning MAD Model~C shows a peculiar set of short timescale,
quasi-regular dimming episodes at $1.3\,\mm$, marked by arrows in the
lightcurves plotted in Figure~\ref{fig:wave}.
The same figure also shows representative images during the flux
minima and recovering phases of two such dimming events.
It is evident that during each dimming event, e.g., the images shown
in the first and third panels from left, the bright funnel region
becomes less active and more collimated.
As the brightness recovers, the funnel becomes active again, with
waves of hotter regions propagating along the funnel walls.
Note that, this effect is quite different from the radio/mm dimming of
\sgra\ discussed in \citet{2010ApJ...724L...9Y}, which may be caused
by obscuration of optically thick plasma.

Among the five best-fit models of \sgra\ that we are considering in
this paper, Model~C is the only one for which the emission is
dominated by the footpoint of the jet funnel \citep[see images
  in][]{2015ApJ...799....1C}.
It is, therefore, not surprising that this is the only model that is
characterized by this dimming behavior.
The individual episodes are short-lived ($\lesssim 1\,\hr$) and
quasi-regular (once every $\sim 1$--$3\,\hr$).
Even though the overall behavior of these events is well captured by
our simulations, the dimming depth is affected by the fact that we
excise the innermost grid cells near the rotational pole of the
spacetime to avoid artificially hot pixels, as documented in
\citet{2015ApJ...799....1C}.

\subsubsection{Magnetic Reconnection}
\label{sec:reconnection}

In the rapidly spinning, jet-dominated MAD Models~D and E, the
relatively large extent of the emission region across the
long-wavelength spectrum and the lack of well defined, short-lived
flaring events makes it difficult to identify the physical origin of
the variability in the simulated lightcurves.
Nevertheless, there is some evidence for variability generated by
dissipation of magnetic energy in localized events, as shown in
Figure~\ref{fig:recon}.

Visual inspection of the images during and after one of these events
reveals that the increase in the emission starts as a localized hot
spot in the X-rays, most probably due to magnetic reconnection and
dissipation of the magnetic energy.
This hot emitting region then spirals upwards, and the change in its
physical location is mirrored in the evolution of its temperature and,
hence, in the peak wavelength of its emission.
This is the origin of the $\sim 1\,\hr$ long time lags between the IR,
$1.3\,\mm$, and radio lightcurves discussed in
section~\ref{sec:correlations}.
This is similar to the coronal mass ejection episodes observed in the
Sun \citep[see, e.g.,][for a review]{2000JGR...10523153F} and is the
closest analogue within GRMHD simulations to the expanding plasmon
model of \citet{1966Natur.211.1131V} that has been used in several
phenomenological description of time lags from \sgra\ \citep[see,
  e.g.,][]{2008ApJ...682..361Y, 2009ApJ...706..348Y,
  2010ApJ...724L...9Y}.

\section{Discussion and Conclusions}

In this paper, we performed an extensive exploration of the predicted
variability properties of GRMHD numerical models of \sgra, with
parameters that have been fixed in earlier work
\citep{2015ApJ...799....1C} by fitting their time-averaged properties
to the observed spectrum of \sgra\ and to the size of its $1.3\,\mm$
image.
In particular, we investigated the variability properties of five
models, two of which were disk-dominated SANE models and three were
jet-dominated MAD models.

We found that the disk-dominated SANE models generated
photon-frequency dependent amplitudes of variability and IR power
spectra that are in agreement with those inferred observationally.
Their lightcurves also show a number of short-lived, large flaring
events that have short ($\lesssim 1\,\hr$) time lags between the IR
and $1.3\,\mm$, but have no X-ray counterparts.
The distribution of IR flare amplitudes is well described by a
log-normal distribution, with potential excess at large amplitudes, in
agreement with the observations.
We showed that strong-field gravitational lensing of hot, magnetically
dominated regions is responsible for these large, achromatic flares.
The reason that such flares were absent in our earlier work
\citep{2009ApJ...701..521C} is because we did not include there the
effects of gravitational lensing and incorporated a more simplified
model for plasma thermodynamics.
However, we also confirmed our earlier result that MHD turbulence
alone with thermal plasma processes does not produce large X-ray
flares, since the X-ray emission originates at large radii, where
gravitational lensing is negligible.

We speculate that particle acceleration processes need to be
incorporated in the plasma model in the GRMHD simulations to account
for the observed X-ray flaring. We expect that the presence of a
short-lived distribution of relativistic electrons in the inner
accretion flow will generate significant flux excursions from the
X-rays down to the IR. However, incorporating such non-thermal
electrons in the model will not affect significantly the properties of
the flares we report here. Indeed, because the large IR flares are the
result of strong-field gravitational lensing of very localized
magnetic filaments, the total scattering optical depth (for Compton
emission) or emission volume (for synchrotron emission) of non-thermal
electrons in these filaments will be too small to generate significant
excursions in the X-ray flux (see also the estimates in
\citealt{2009ApJ...698..676D}, section~6).
In contrast to the SANE models, the variability of the MAD models
shows only slow trends, with no evidence for short-lived flares at any
of the wavelengths.
The non-spinning jet-dominated model, where the $1.3\,\mm$ emission is
localized at the footpoint of the funnel, shows quasi-regular dimming
events that arise from the time-dependent loading of the funnel.
The rapidly spinning jet models show weak evidence for time lags
between wavelengths, with longer wavelengths lagging the shorter ones.
These time lags seem to arise from the upward motion of hot regions
heated by magnetic reconnection/dissipation events.
Overall, the variability properties of the MAD simulations do not
appear to agree with those inferred observationally for \sgra.

Our results indicate that, in the near future, it will be possible to
observe strong-field gravitational lensing effects around black holes
with the Event Horizon Telescope at $1.3\,\mm$ and with GRAVITY in the
IR.
Even though the flare events are too short-lived to obtain full images
during a single lensing event, their presence can be inferred from the
evolution of the closure-phases obtained along triangles with
appropriate baselines \citep{2009ApJ...695...59D,
  2014MNRAS.441.3477V}.

\acknowledgements

C.K.C., D.P., and F.O. were partially supported by NASA/NSF TCAN award
NNX14AB48G and NSF grants AST~1108753 and AST~1312034.
D.M. acknowledges support from NSF grant AST-1207752.
R.N. acknowledges support from NASA/NSF TCAN awards NNX14AB47G.
All ray tracing calculations were performed with the \texttt{El Gato}
GPU cluster at the University of Arizona that is funded by NSF award
1228509. We thank Jon McKinney for valuable comments and discussions.

\bibliography{my,ms}

\end{document}